\documentclass[%
 reprint,
 amsmath, amssymb,
prstab,
]{revtex4-1}

\usepackage{graphicx}
\usepackage{dcolumn}
\usepackage{bm}

\begin{document}

\preprint{APS/123-QED}

\title{Evaluation of Beam Halo from Beam-Gas Scattering at the KEK-ATF}
\thanks{ryang@lal.in2p3.fr}%

\author{R. Yang$^{1, *}$, T. Naito$^{2, 3}$, S. Bai$^{4}$,  A. Aryshev$^{2,3}$, K. Kubo$^{2,3}$, T. Okugi$^{2,3}$, N. Terunuma$^{2,3}$, D. Zhou$^{2,3}$, A. Faus-Golfe$^{1}$, V. Kubytskyi$^{1}$, S. Liu$^{5}$, S. Wallon$^{1}$}
 
\author{P. Bambade$^1$}
\affiliation{
$^{1}$LAL, Univ. Paris-Sud, CNRS/IN2P3, Universit$\acute{e}$ Paris-Saclay, Orsay, France\\
$^{2}$High Energy Accelerator Research Organization,  Tsukuba, Ibaraki, Japan\\
$^{3}$School of High Energy Accelerator Science, SOKENDAI, Tsukuba, Ibaraki, Japan \\
$^{4}$Institute of High Energy Physics,  Beijing, China\\
$^{5}$Deutsches Elektronen-Synchrotron, Hamburg, Germany
}%
%
%
%

\date{\today}

\begin{abstract}

In circular colliders, as well as in damping rings and synchrotron radiation light sources, beam halo is one of the critical issues limiting the performance as well as potentially causing component damage and activation. It is imperative to clearly understand the mechanisms that lead to halo formation and to test the available theoretical models.  Elastic beam-gas scattering can drive particles to large oscillation amplitudes and be a potential source of beam halo.  In this paper, numerical estimation and Monte Carlo simulations of this process at the ATF of KEK are presented. Experimental measurements of beam halo in the ATF2 beam line using a diamond sensor detector are also described, which clearly demonstrates the influence of the beam-gas scattering process on the transverse halo distribution.

\begin{description}
\item[DOI]

\end{description}
\end{abstract}

\pacs{11}
\maketitle


\section{\label{sec:level1}INTRODUCTION}
\indent In high energy lepton colliders, the balance between the requirements of high luminosity and low detector backgrounds is always a struggle. To control the background induced by halo particles with large betatron amplitude or energy deviation,  a robust collimation system upstream is essential. The design of collimators requires some knowledge of the halo distribution and population, to estimate the collimation efficiency~\citep{stancari2011collimation}. To describe the halo distribution and mechanisms for its formation, a number of numerical and experimental investigations have been performed, for both circular and linear machines~\citep{allen2002beam, tenenbaum2001sources, qiang2000beam, wangler2004beam, wittenburg2009beam}. These studies indicate that halo distributions are influenced by many factors, e.g., space charge, scattering (elastic and inelastic beam-gas scattering, intra-beam scattering and e$^-$cloud), optical mismatch, chromaticity and other optical aberrations, and so on. Moreover, the dominant halo source might be different for each machine, depending on its design and status.  

\indent For the future linear colliders, it is essential to determine plausible halo distributions at the entrance of the main linac and their physical origin. 
The Accelerator Test Facility (ATF) of KEK, which has successfully achieved small emittances satisfying the requirements of the International Linear Collider (ILC), and which includes an extraction line (ATF2) capable of focusing the beam down to a few tens of nanometers at the virtual Interaction Point (IP),  is an ideal machine to study halo formation mechanisms and develop the specialized instrumentation needed for the measurements.  At the ATF2 beam line, the reduction of the modulation in the beam size measurement using the \textit{Shintake} monitor~\citep{yan2014measurement} at the IP due to halo particle loss upstream also motivates a good understanding of halo formation and  ways to suppress it. Considerable efforts have been devoted to reveal the primary mechanism controlling halo formation at ATF~\citep{hirata1992nongaussian, sliuPhD, naito2016beam, dou2014analytical}. The theory to characterize beam profile diffusion due to elastic beam-gas scattering (BGS) has been developed, but has not yet been fully validated experimentally, mainly due to the lack of appropriate instrumentation with high enough dynamic range (DNR,  $\geq10^5$). To achieve a suffcient DNR, a set of diamond sensor detectors (DS) has been constructed and installed at the end of the ATF2 beam line ~\cite{liu2016vacuum}.

\indent  In this paper, numerical evaluations of beam halo from BGS are described, followed by a detailed simulation of halo formation in the presence of radiation damping, quantum excitation, residual dispersion, $xy$ coupling and BGS in damping ring. Halo measurements using the diamond sensor detector are described , which confirm that the vertical halo is dominated by BGS. The results are then discussed and some conclusions and further work are outlined.
 
\subsection{\label{sec:level2}Accelerator Test Facility 2}
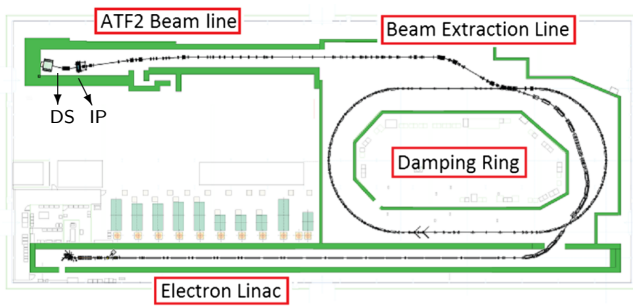
\begin{figure} [htpb]
		\centering
		\includegraphics[width=\linewidth]{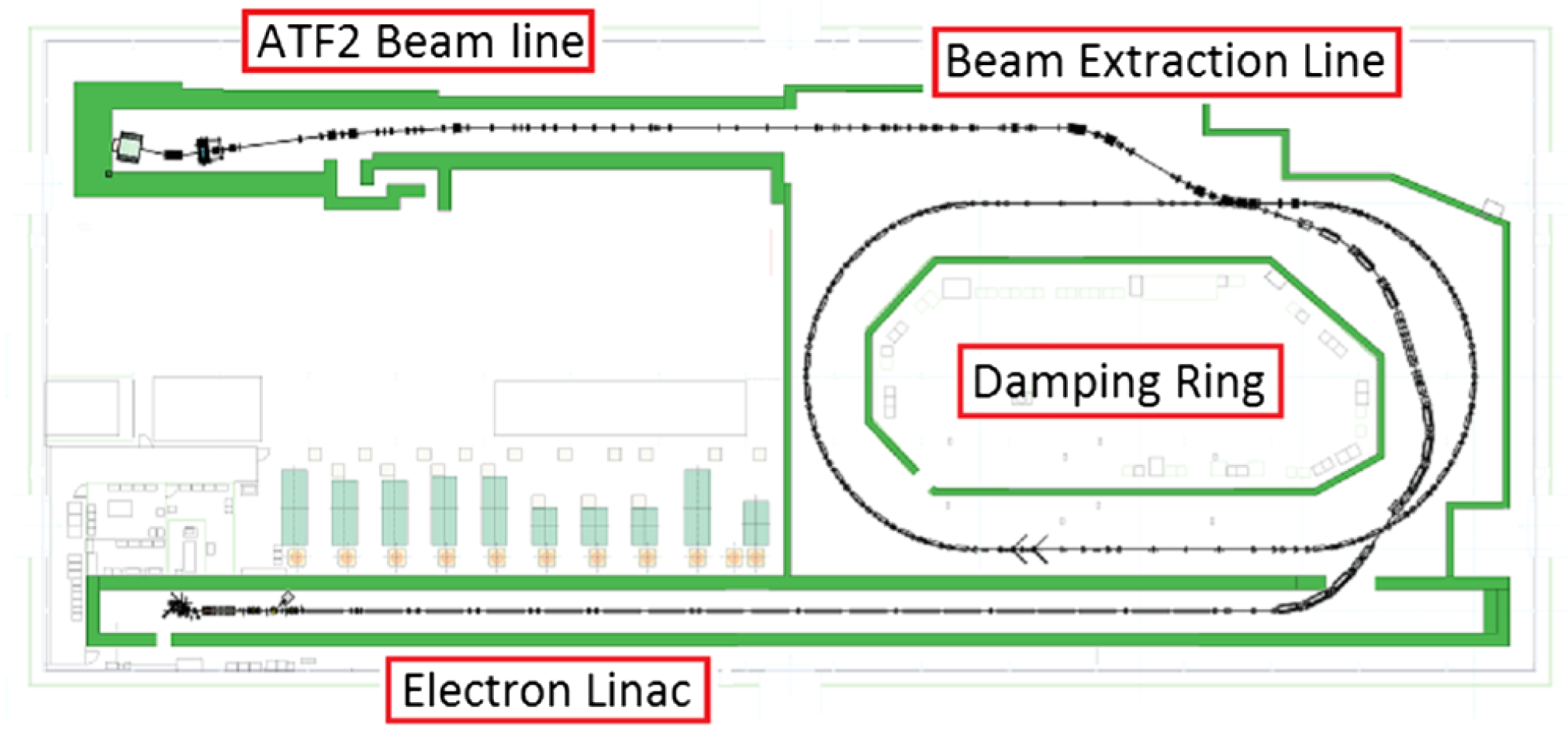}
		\put(-214, 87) {\vector(1, -2){5}}
		\put (-210, 70) {\scriptsize \textsf{IP}}
		\put (-222, 89) {\vector(0, -1) {12}}
		\put(-225, 70) {\scriptsize \textsf{DS}}
		\caption{ Schematic diagram of ATF linac, damping ring and ATF2 beam line, taken from \cite{okugi2014linear}}
		\label{fig:atf2layout}
	\end{figure}

\indent ATF consists of a 1.3 GeV $S-$band linac, a damping ring  and an extraction line, as shown in Fig.~\ref{fig:atf2layout}. The smallest vertical rms emittance measured at low intensity was 4 pm~\cite{honda2004achievement}, which corresponds to the normalized emittance of 1.1$\times10^{-8}$ m. The main beam parameters in the ATF damping ring are summarized in Table~\ref{tab:table1}.
	
\indent As an extension of ATF, ATF2 aims to address the feasibility of focusing the beam to a few tens of nanometer size and providing the beam orbit stabilization of the nanometer level at the IP.  ATF2 is also an energy-scaled version of the compact focusing optics designed for the ILC, using the local chromaticity correction scheme~\cite{raimondi2001novel, white2014experimental}. 
	\begin{table} [htb] 
	\caption{\label{tab:table1} ATF main parameters\citep{hinode1995atf, PhysRevLett.88.194801}}
	\begin{ruledtabular}
			\begin{tabular}{ccc}
			Beam energy [GeV] &	$E_0$ &	1.3 \\
			Intensity [$e$/pulse] & $N$	&  1-10$\times10^{9}$\\
			Vertical emittance [pm] & $\epsilon_y$	& $>$4	\\
			Horizontal emittance [nm]	& $\epsilon_x$	&	1.2	\\
			Energy spread [\%]	&	$\sigma_\delta$	&	0.056 (0.08)\footnote{with intra-beam scattering (IBS) for the beam intensity of $1\times10^{10}e$/pulse}	\\
			Bunch length [mm]	&	$\sigma_z$	&	5.3	(7)$^\text{a}$\\
			Damping time [ms]	&	$\tau_x$/$\tau_y$/$\tau_z$	&	17/27/20\\
			Injection emittance [nm] & $\epsilon_{x0}$/$\epsilon_{y0}$ & 14 \\
			Storage time[ms] & $t$ & 200 \\
			Momentum acceptance [\%] & $\Delta p/p$ & 1.2 \\			
			\end{tabular}	
	\end{ruledtabular}
	\end{table}
	
\section{Theoretical Evaluation}
\subsection{Analytic Approximation}
\indent We follow the approach developed by K. Hirata~\cite{hirata1992nongaussian} for the description of particles' redistribution in the presence of stochastic processes. The transverse motion in a ring or transport beam line can be perturbed by stochastic processes such as synchrotron radiation, BGS or IBS. It can be described by the diffusion equation 
	\begin{equation}
		\frac{d\vec{x}}{ds} = -[H(\vec{x},s), \vec{x}]+\xi(\vec{x},s)
	\end{equation}
where $\vec{x}$ is the 6D phase  space coordinate, $H(\vec{x},s)$ the Hamiltonian representing the symplectic part of the motion and $\xi(\vec{x},s)$ contains the diffusion effects. The solution to the equation of motion can be expressed in terms of a  linear map and the integrated perturbation of the stochastic process
	\begin{equation}\label{eq:transfMap}
		\vec{x}(s) = M(s, s_0)\vec{x}_0+\int_{s_0}^{s}M(s,s')\tilde{\xi}(s)ds'
	\end{equation}
with
	\begin{equation}\label{eq:transfMatrx}
		M(s, s_0) = M_0\exp[\int_{s_0}^{s}[s\tilde{H}(s'')-D(s'')]ds'']
	\end{equation}
where $M_0$ is the symplectic matrix representing the linear transformation, $\tilde{H}$ a symmetric 6$\times6$ matrix and $D$ the damping matrix which contains the radiation damping~\cite{ohmi1994beam}. Here we describe only the transverse motion (in the horizontal plane for example) and we consider only the betatron motion, radiation damping, quantum excitation and diffusion from BGS, ignoring betatron coupling. In the normalized coordinates $u=x/\sqrt{\beta}$ and $u'=du/d\phi$, Eq. (\ref{eq:transfMap}) can be written as
	\begin{equation}\label{eq:diffusionEq0}
		\centering	
		\vec{u}(s) = R(s,s_0) \vec{u}(s_0) \exp(-\frac{\alpha }{c_0}\int_{s_0}^{s}ds)+\delta \vec{u}
	\end{equation}
where $\vec{u}=(u, u')^T$, $R(s, s_0)$ is a pure rotation, $\alpha$ the damping rate and  $\delta \vec{u}$ the perturbation, expressed as
	\begin{eqnarray}
		R(s, s_0) &&= \begin{pmatrix}
						\cos(\Delta\phi) && \sin(\Delta\phi) \\
						-\sin(\Delta\phi) && \cos(\Delta\phi)
					\end{pmatrix}\\
		\delta \vec{u} &&= R(s,s_0) \begin{pmatrix}
								0 \\
								\sqrt{\beta}\theta_x 
							\end{pmatrix} \text{exp}(-\frac{\alpha }{c_0}\int_{s_0}^{s}ds)		
	\end{eqnarray}
where $\Delta\phi=\int_{s_0}^{s}\frac{ds}{\beta(s)}$ is the phase advance, $\beta$ the betatron function, $\theta_x$ the transverse kick angle at $s_0$ and $c_0$ the light velocity in vacuum. We can further specify the perturbation term in Eq.(\ref{eq:diffusionEq0}) in terms of the transformation in presence of radiation damping, diffusion due to the quantum excitation, $\delta\vec{u}_{qe}$ and to the external perturbation due to BGS, $\delta\vec{u}_{ex}$ 
	\begin{equation} \label{eq:diffusionEq1}
		\centering
		\vec{u}(s) = R(s, s_0) \vec{u}(s_0) \exp(-\frac{\alpha}{c_0}\int_{s_0}^sds) + \delta\vec{u}_{qe} + \delta\vec{u}_{ex}		
	\end{equation}
	
The stationary distribution is determined by the integral of all stochastic processes. Since particle distribution under the influence of radiation damping and quantum excitation has been well understood, it is convenient to express the distribution function $\psi(u)$ as
	\begin{equation}\label{eq:distribfuncdiffu}
		\psi(u) = \frac{1}{2\pi}\int e^{i\omega u} \tilde{\psi}_t(\omega) \tilde{\psi}_f(\omega)d\omega
	\end{equation}
where $\tilde{\psi}_t(\omega)$ is the characteristic function in presence of radiation damping and quantum excitation 
	\begin{equation}
		\tilde{\psi}_t(\omega)=\exp(-\omega^2\sigma_{t}^2/2)
	\end{equation}
in which $\sigma_t$ is the beam size in absence of external perturbation. The characteristic function $\tilde{\psi}_f(\omega)$ has been derived in Ref.~\cite{hirata1992nongaussian} and Ref.~\cite{raubenheimer1992emittance}, thanks to Campbell's theorem~\cite{rice1944mathematical}. Here, we use the formalism in Ref.~\cite{hirata1992nongaussian} where the stochastic perturbation is treated over many betatron oscillation periods . Approximating $\beta$ by its average  over the ring, $\bar{\beta}$, the characteristic function $\tilde{\psi}_f(\omega)$ can be written as
	\begin{equation}
		\tilde{\psi}_f(\omega)  = \exp(\frac{N}{\alpha}\hat{f}(\omega\sqrt{\bar{\beta}}))	
	\end{equation}
	
	\begin{equation}\label{eq:diffusiontermFre}
		\hat{f}({\tilde{\omega}})  = \frac{2}{\pi}\int_{0}^{1} d\zeta \frac{\Re[\tilde{f}(\tilde{\omega}\zeta)]-1} {\zeta} \cos^{-1} \zeta
	\end{equation}
	
	\begin{equation}
		\tilde{f}(\tilde{\omega})  = \int d\theta_x f(\theta_x) \cos(\tilde{\omega}\theta_x)
	\end{equation}
where $N$ is the scattering rate of a test particle, $\Re[\tilde{f}(\tilde{\omega}\zeta)]$  the real part of  $\tilde{f}(\tilde{\omega}\zeta)$ and $f(\theta_x)$ is the probability distribution of the deflection angle $\theta_x$.  Then, the final distribution function can be expressed as
\begin{equation} \label{eq:disfunc_bgs_final0}
	\centering	
	\psi(u) = \frac{1}{2\pi}\int_{-\infty}^{\infty}e^{i\omega u}\exp(-\frac{\omega^2\sigma_t^2}{2} + \frac{N}{\alpha}\hat{f}(\omega\sqrt{\bar{\beta}})) d\omega
\end{equation}
The characteristic function is an even function, only the cosine part remains after performing the integration. The distribution can be further expressed as
\begin{equation} \label{eq:disfunc_bgs_final1}
	\centering
	\psi(u) = \frac{1}{\pi}\int_{0}^{\infty}\cos(\omega u)\exp(-\frac{\omega^2\sigma_t^2}{2} + \frac{N}{\alpha}\hat{f}(\omega\sqrt{\bar{\beta}})) d\omega
\end{equation}
Moreover, the transverse distribution in $x$ can be described by
\begin{equation} \label{eq:disfuc_bgs_final2}
	\centering
	\psi(x_i) = \frac{1}{\pi}\int_{0}^{\infty}\cos(\omega x_i)\exp(-\frac{\omega^2\sigma_{x_i}^2}{2} + \frac{N}{\alpha}\hat{f}(\omega\sqrt{\bar{\beta}\beta_i})) d\omega
\end{equation}
where $x_i$ is the horizontal coordinate at the $i$-th element, $\sigma_{x_i}$  the equilibrium horizontal beam size in presence of radiation damping and quantum excitation, and $\beta_i$ the beta function at the observation point.

\indent To obtain the numerical form of the distribution function $\psi(u)$ or $\psi(x)$, we have to firstly evaluate the $\tilde{f}(\tilde{\omega})$ function in the presence of BGS. Treating BGS as the classical Rutherford scattering process, the cross section in the cgs. units is given by
\begin{equation} \label{eq:cross-secBGS}
	\centering
	\frac{d\sigma}{d\Omega} = (\frac{2Zr_e}{\gamma})^2\frac{1}{(\theta^2 +\theta_m^2)^2}
\end{equation}
where $\Omega$ is the solid angle, $Z$ the atomic number, $r_e$ the classical electron radius, $\gamma$ the Lorentz factor, $\theta$ the transverse deflection angle and $\theta_m$ the minimum deflection angle determined by the uncertainty principle.
\begin{equation} \label{eq:theta_m}
	\centering
	\theta_m = \frac{Z^{1/3}}{\alpha_0\gamma}
\end{equation}
where $\alpha_0$ is the fine structure constant. The transverse deflection angle $\theta$ can be further specified as
\begin{equation}
	\centering
	\theta^2 = \theta_x^2 + \theta_y^2
\end{equation}
Note $\theta_x\in[-\theta_{x,\text{max}}, \theta_{x, \text{max}}]$ and the same for $\theta_y$. The $d\sigma/d\theta_x$ can be obtained by integration of Eq.(\ref{eq:cross-secBGS}) over vertical deflection angle $\theta_y$. If we assume $\theta_{y, \text{max}}\gg\sqrt{\theta_x^2+\theta_m^2}$, $d\sigma/d\theta_x$  can be approximated as
\begin{equation} \label{eq:BGScrossapprox}
	\centering
	\frac{d\sigma}{d\theta_x} \approx \frac{\pi}{2}(\frac{2Zr_e}{\gamma})^2\frac{1}{(\theta_x^2 + \theta_m^2)^{3/2}}
\end{equation}
Then the total cross section $\sigma_{tot}$,  probability function $f(\theta_x)$ and scattering rate $N$ becomes
\begin{align} \label{eq:cross-sec-total}
	\centering
	\sigma_{tot} & = \int_{-\theta_{x, \text{max}}}^{\theta_{x, \text{max}}}\frac{d\sigma}{d\theta_x}d\theta_x =  \frac{4\pi Z^2r_e^2}{\gamma^2\theta_m^2} \\
	 f(\theta_x) & = \frac{1}{\sigma_{tot}}\frac{d\sigma}{d\theta_x} = \frac{\theta_m^2}{2(\theta_x^2+\theta_m^2)^{3/2}} \\
	N &= \rho_v\sigma_{tot}c_0		 
\end{align}
where $\rho_v$ is the volume density of residual gas atoms. Following the derivation in Ref. \cite{hirata1992nongaussian}, function $\tilde{f}(\tilde{\omega})$ and $\hat{f}(\tilde{\omega})$ are finally expressed as
\begin{equation}
	\begin{split}
	\centering
	\tilde{f}(\tilde{\omega}) & = \tilde{\omega}J_1(\tilde{\omega})\\
	\hat{f}(\tilde{\omega}) & = \frac{2}{\pi}\int_0^1d\zeta\frac{\tilde{\omega}J_1(\tilde{\omega})-1}{\zeta}\cos^{-1}\zeta
	\end{split}
\end{equation}
where $J_1$ is the modified Bessel function of the first order. Estimations of beam profiles  using Eq.(\ref{eq:disfuc_bgs_final2}) for the ATF damping ring are later shown in Fig.\ref{fig:simulationTheo}.
\subsection{Tracking Simulation}
\begin{figure}  [htbp]
			\centering
			\includegraphics[width=\linewidth]{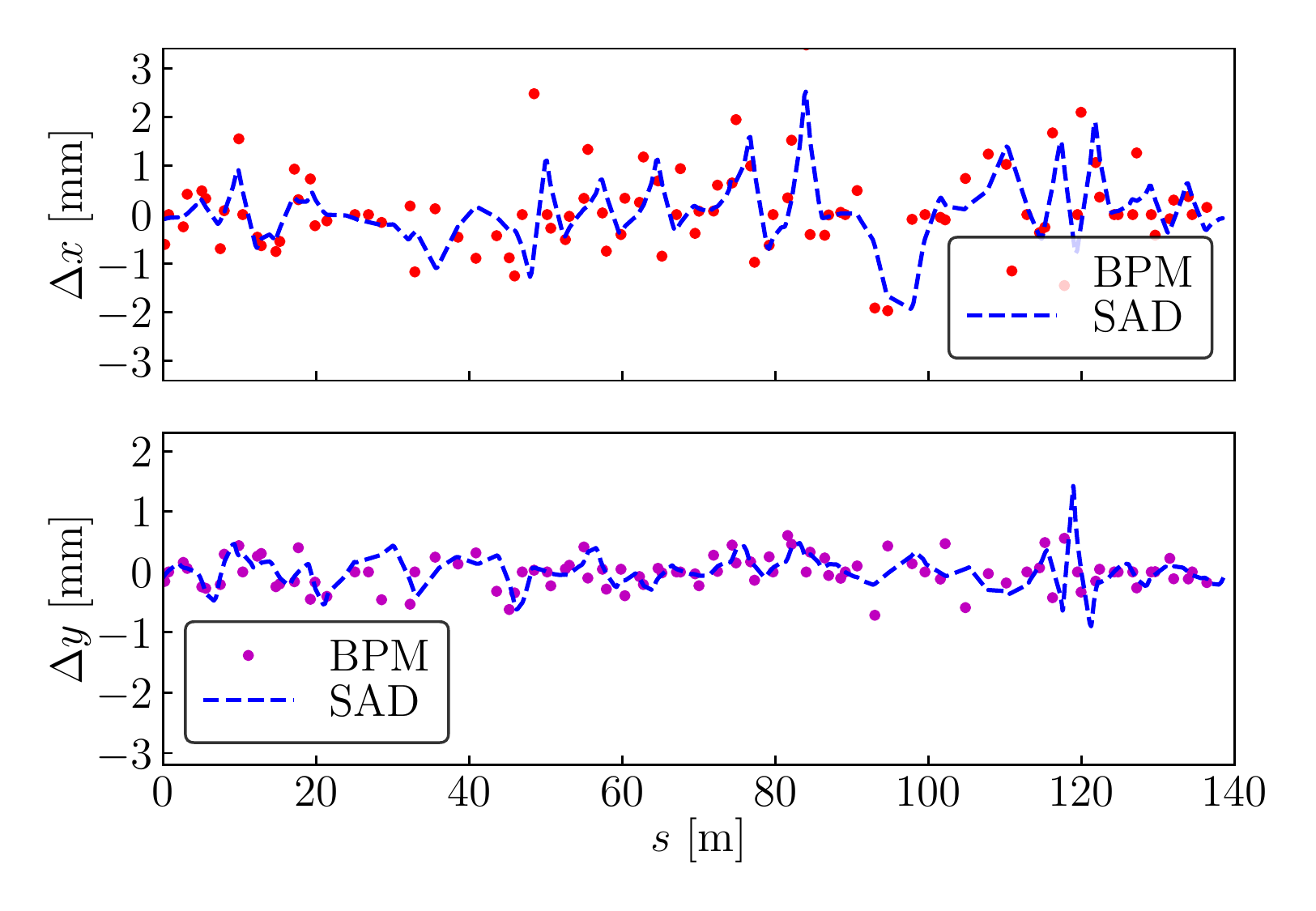}
			\put(-200, 148){(a)}
			\put(-200, 76){(b)}
			\caption{Horizontal (a) and vertical (b) COD measured by BPMs in January 2017 and approached by local orbit bumps}
			\label{fig:dr_exeydxdy}
		\end{figure}

\indent Generation and tracking of core particles and scattered particles are performed through a script developed based on SAD~\cite{SAD}, a program  used for optical matching and closed orbit distortion (COD) correction during beam operation. The equilibrium vertical emittance $\epsilon_y$ is mainly determined by the residual vertical dispersion $\eta_y$ and cross-plane betatron coupling, both of which strongly depend on the residual alignment errors of magnets and COD~\citep{hinode1995atf, kubo2003simulation}. The observed vertical emittance can be approached by introducing random effective vertical displacements to quadrupoles and sextupoles ($20 \mu$m, RMS),  respectively, and rotations of quadrupoles (2 mrad, RMS). The equilibrium emittance $\epsilon_y$, obtained for various seeds, ranges from 5 pm to 30 pm.  Alternatively, the actual COD, which is measured by BPMs,  can also be approached by local orbit bumps using steering magnets, as shown in Fig.~\ref{fig:dr_exeydxdy}. Equilibrium emittances are 12 pm and 1.2 nm, vertically and horizontally, respectively, for a realistic COD.  The latter can better represent the realistic orbit and beam parameters, and is therefore used in our BGS simulations. The emittances and beam sizes considered here and in the following are evaluated by Gaussian fits to the beam core distributions.

\indent Tracking of both scattered and non-scattered particles is performed element-by-element, separately, utilizing the common beam parameters at injection, as shown in Table~\ref{tab:table1}. The simulation of scattered particles is performed as follows~\cite{ryang2017iop}. At first, in each turn, the number of scattering events and their perturbations are predicted randomly according to the residual gas pressure and the cross section. Secondly, perturbations in the 6D phase space of particles are implemented at random longitudinal positions to generate the scattered particles. The location of scattered particles is approximated to be at the closest element, which determines the local Twiss parameters and orbit. Thirdly, the scattered particles in the present turn are transported to the observation point (at the location of the extraction kicker), to be combined with the scattered particles accumulated from the previous turns. The above process is then repeated until beam extraction. In addition, the possibility of multi-BGS has also been considered. 

\indent In order to estimate beam profiles 
 in the ATF2 beam line, stored particles are then extracted and transported to diagnostic points. Initial Twiss parameters of the ATF2 lattice are well matched with the DR lattice at the extraction kicker. Orbit distortion of the extracted beam in the kicker-septum region is represented by the coordinate transformation.  The "10$\times$1" optics~\cite{okugi2014linear} of the ATF2, with beta-functions of $\beta_x=40$ mm and $\beta_y=0.1$ mm at the IP, is used.

\indent Estimation of the vacuum lifetime $\tau_v$, which depends directly on the gas pressure, supplies a benchmark for the simulation.  
We assume Z = $\sqrt{50}$ and two atoms per molecule, which approximates air or CO~\citep{hirata1992nongaussian}, to represent the residual gas . For the averaged gas pressure of $1\times10^{-6}$ Pa, the calculated value of $\tau_v$ is 83 minutes~\cite{lifetime_formula}. Meanwhile, the simulated value is 87 minutes, with the equilibrium beam parameter and realistic physical apertures.

\begin{figure} [h!]
	\includegraphics[width=\linewidth]
	{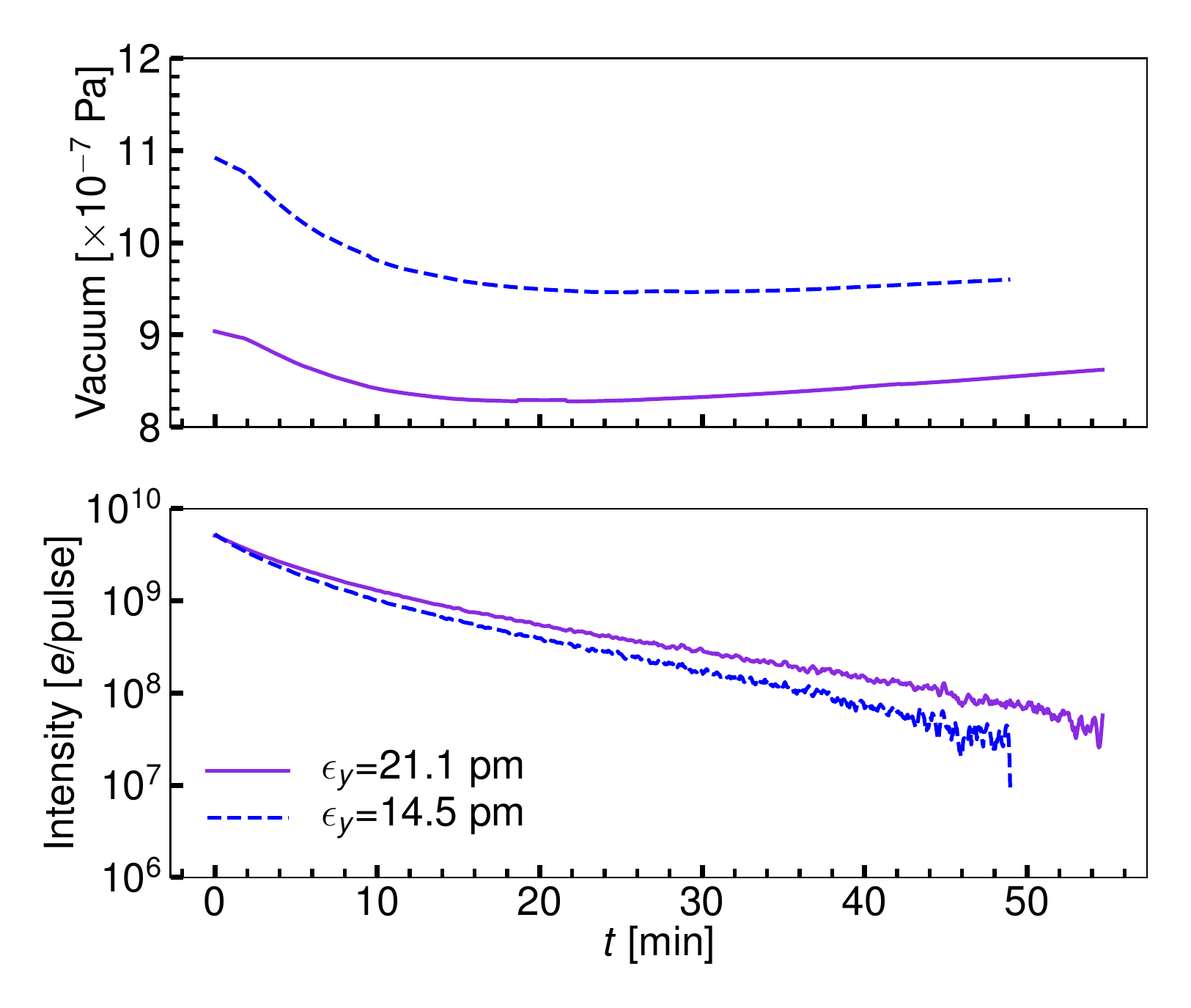}
	\put(-33, 177){(a)}
	\put(-33, 82){(b)}
	\caption{ Evolution of the averaged gas pressure (a) and current decay  of the stored beam (b) in the ATF damping ring}
	\label{fig:VacuLifetime}
	\end{figure}

\indent Vacuum lifetime was also measured at ATF, assuming beam lifetime is dominated by Touschek scattering and elastic BGS. The time dependence of the beam intensity can be described by
\begin{equation} \label{eq:dr_lifetime}
	\centering
	n(t) = 1 - \alpha\int_{0}^{t}dt'P(t')n(t') - \frac{1}{\tau_{Tou}(\kappa)}\int_{0}^{t}n^2(t')dt'
\end{equation}
where $n(t)=N(t)/N_0$ is the normalized beam intensity, $\alpha=1/(\tau_v P)$  a coefficient related to the vacuum lifetime $\tau_v$ and gas pressure $P$ and $\tau_{Tou}$ is the Touschek lifetime.
The decay of the beam current and the variation of the average gas pressure are shown in Fig.~\ref{fig:VacuLifetime} for different vertical emittances.  The coefficient $\alpha$ is around 1000 Pa$^{-1}\cdot$s$^{-1}$, and $\tau_v\approx16$ minutes,  as determined by fitting the current decay with Eq.(\ref{eq:dr_lifetime}). Such a reduction in the experimentally measured vacuum lifetime has been reported in Ref.~\cite{zimmermann1998measurements} and Ref.~\cite{okugi2000evaluation}, which suggested the probable reasons might be: 1) existence of a larger horizontal beam halo induced by other mechanisms; 2) reduction of the dynamic aperture due to sextupole components at the entrance/exit of the combined function bending magnets.
	
\begin{figure} [htp]
	\includegraphics[width=0.9\linewidth]{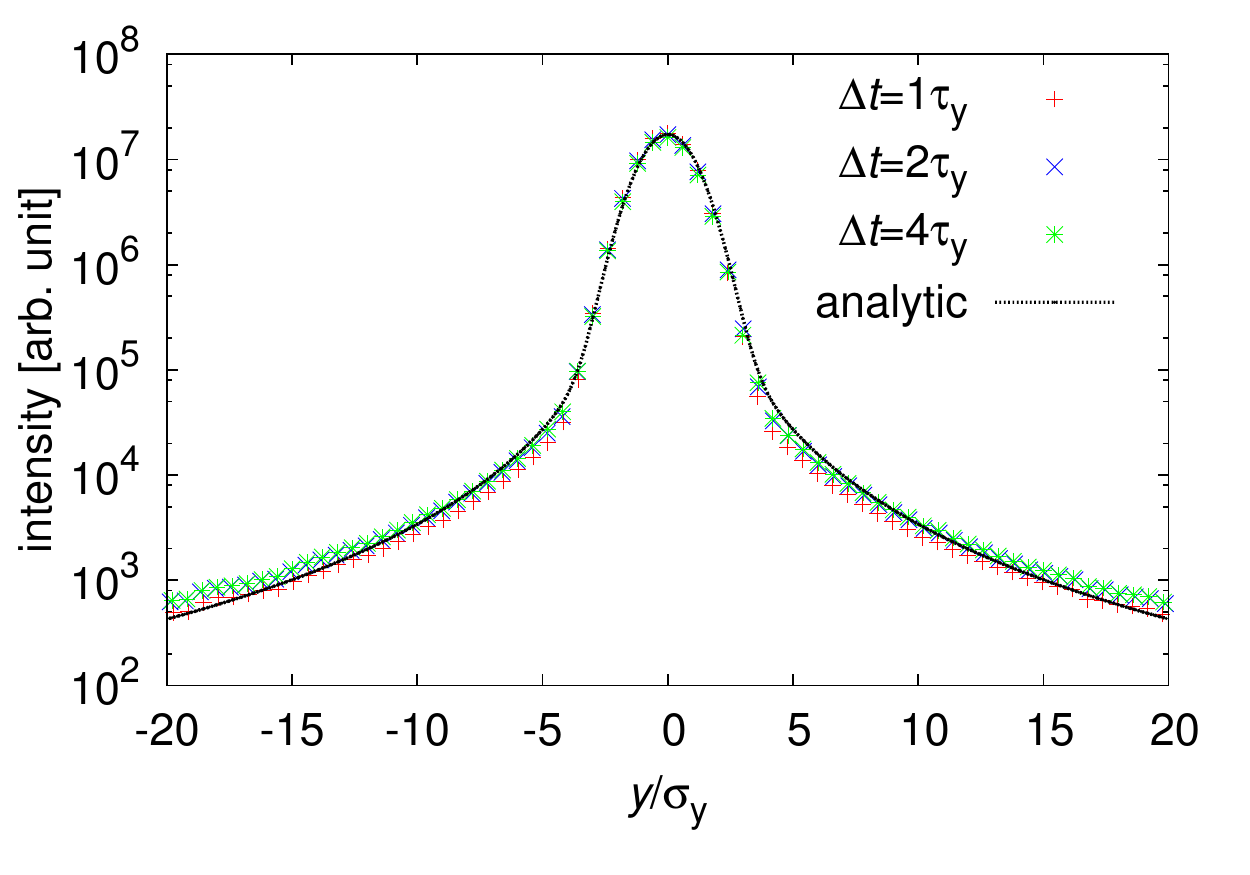}
	\put(-180, 130){(a)}\\*
	\includegraphics[width=0.9\linewidth]{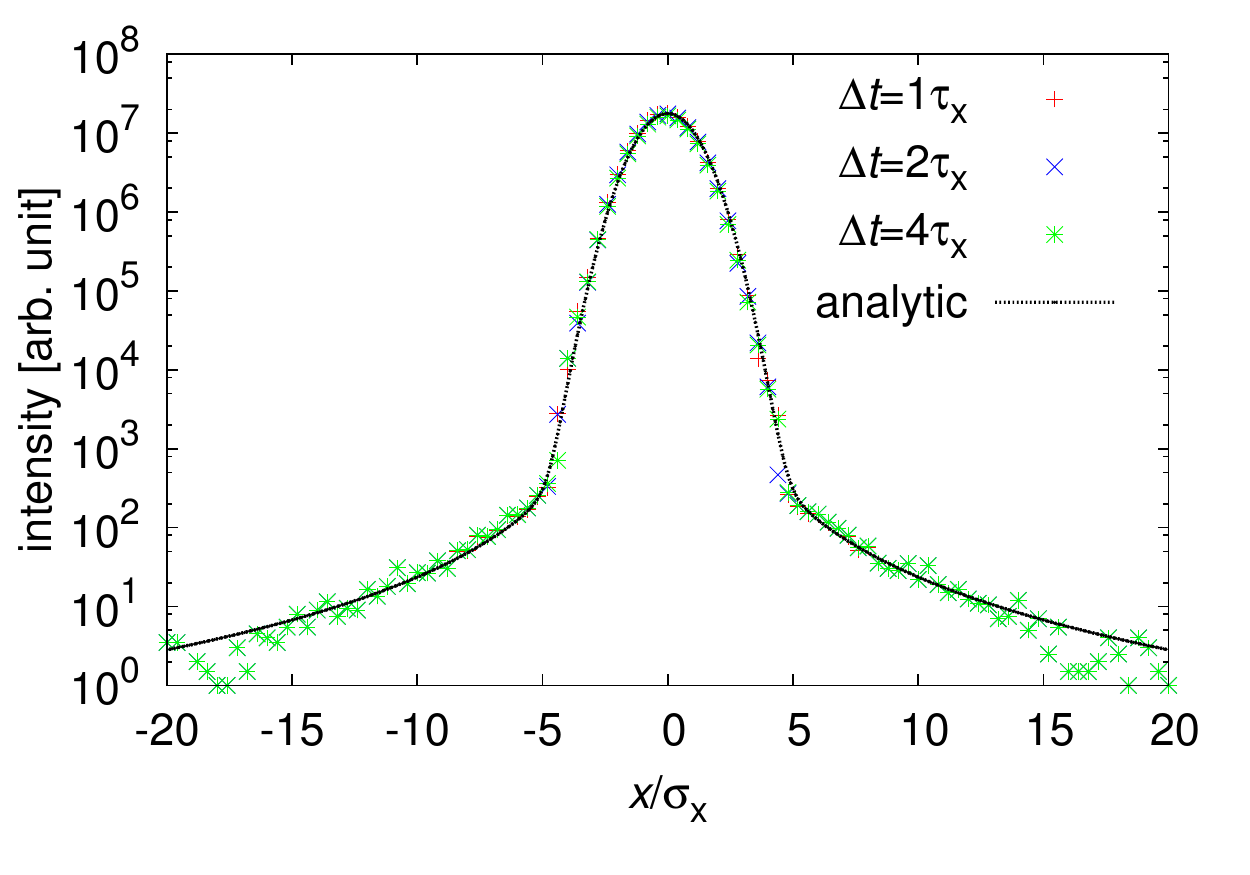}
	\put(-180, 130){(b)}
	\caption{\label{fig:simulationTheo} Comparison of vertical (a) and horizontal (b) beam distortion between analytic approximation and simulation. A tracking time of more than 2 damping times is essential to reach the equilibrium. }
	\end{figure}

\indent 	 The cross section of elastic beam-gas scattering is inversely proportional to $\theta^2$ and therefore large angle events are quite infrequent. Thus, we set an upper bound on the scattering angle at 100 $\theta_m$, which is much larger than the RMS divergences of core particles. The minimum deflection angle $\theta_m$ for ATF beam is 5.6 $\mu$rad. To acquire sufficient statistics, the number of accumulated scattering particles can be as many as 2$\times10^7$. These simulations firstly indicate that at least twice the damping time is essential to reach equilibrium redistribution in the ATF damping ring. For the typical vacuum level of 5$\times10^{-7}$ Pa, satisfactory agreement between the analytical calculation using Eq. (\ref{eq:disfuc_bgs_final2}) and the simulation is observed, see Fig.~\ref{fig:simulationTheo}, where the distribution is normalized to the core beam size. After such a normalization, the horizontal tail/halo appears lower than the vertical halo by around two orders of magnitude, due to the flat aspect ratio of the ATF beam, the horizontal beam size being typically ten times larger than the vertical.
			
\begin{figure} [htp]
	\includegraphics[width=0.9\linewidth]{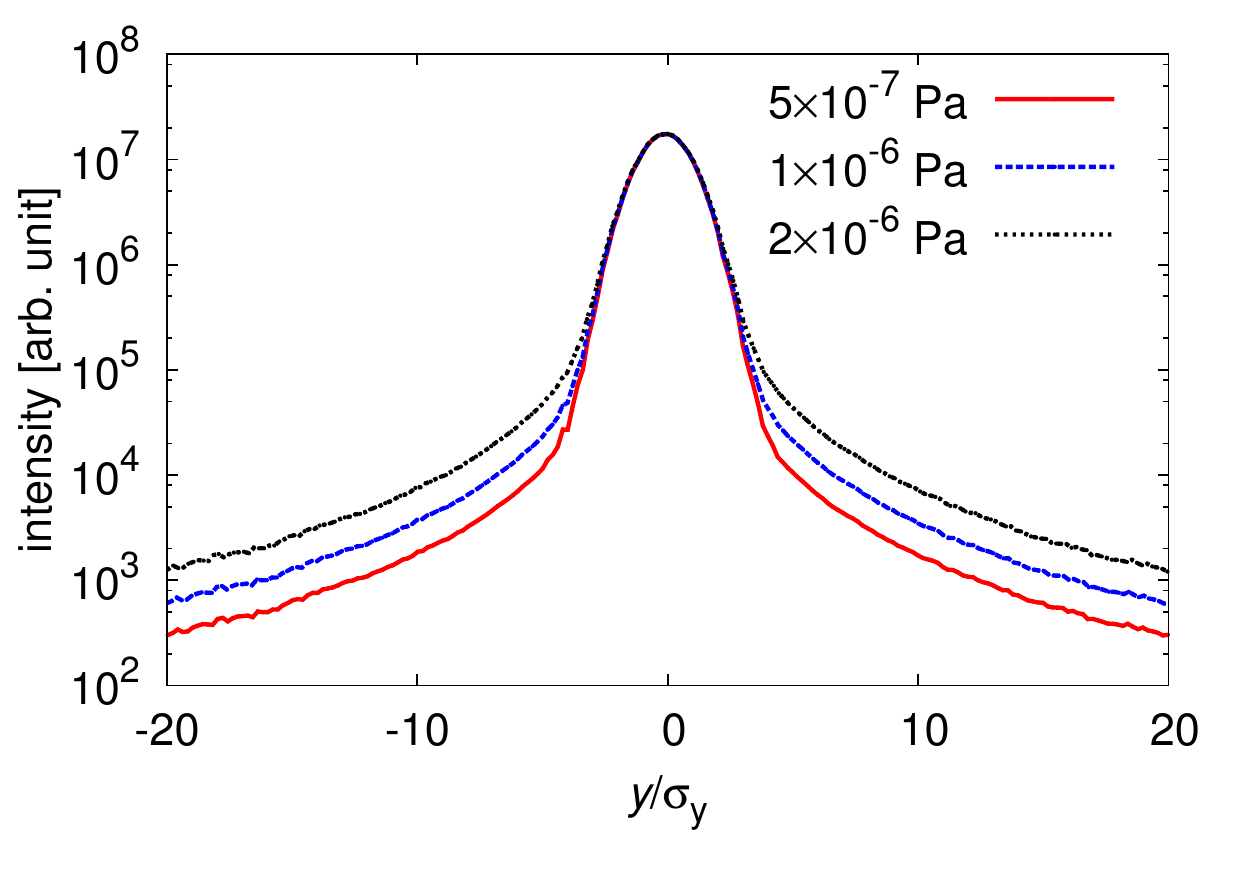}
	\put(-180, 130){(a)}\\*
	\includegraphics[width=0.9\linewidth]{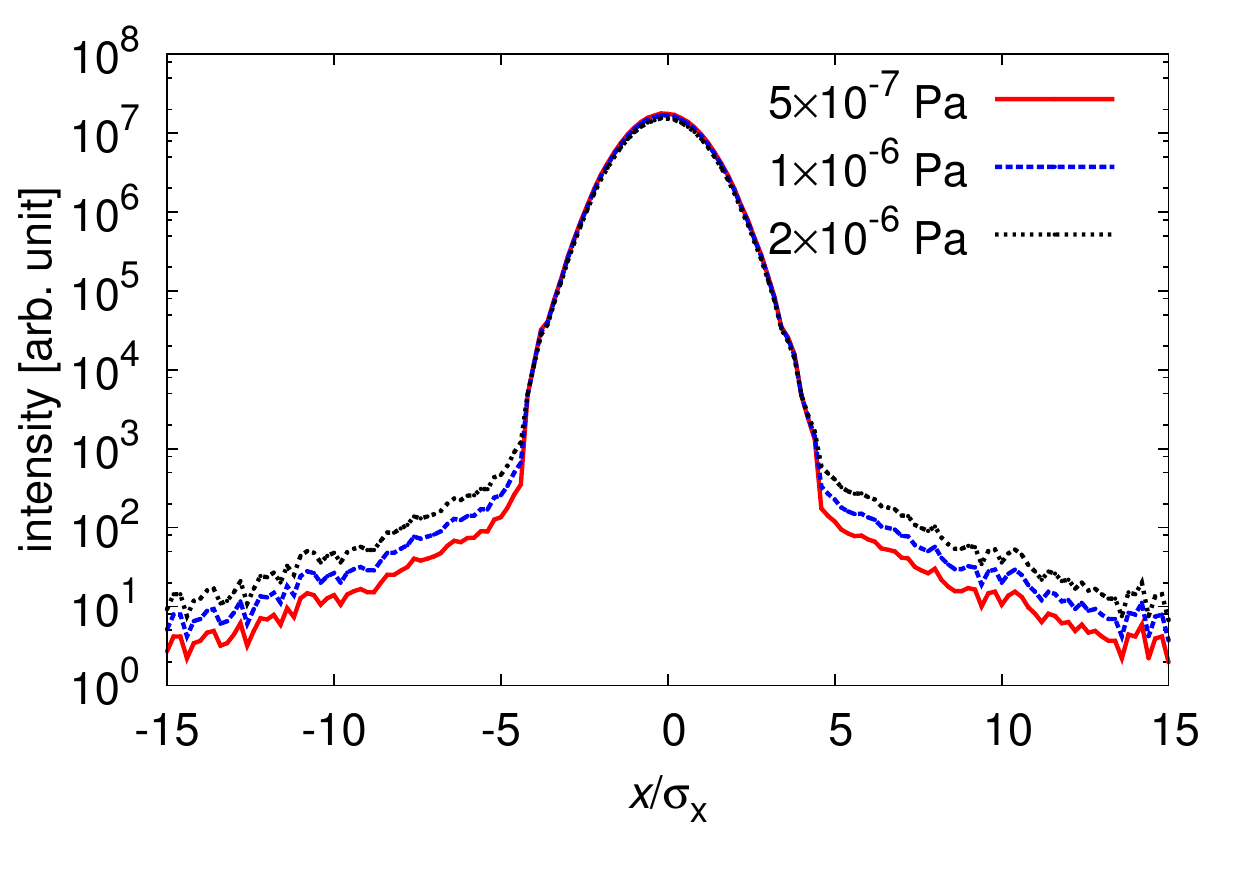}
	\put(-180, 130){(b)}
	\caption{\label{fig:simulationVacuDep} Vacuum dependence of vertical (a) and horizontal (b) beam profiles}
	\end{figure}

\indent 	The probability of BGS depends on the density of the residual molecules, and therefore, beam halo will be increased for a worsened vacuum pressure in the ring. Presently, the averaged gas pressure obtained in the normal operation is $2\times10^{-7}$ Pa, which can be adjusted by turning off part of the sputtering ions pumps (SIPs). Simulations have been performed for three different vacuum levels which were achieved in operation. Significant increases of the beam tail/halo can be observed for the worsened vacuum conditions, as shown in Fig.~\ref{fig:simulationVacuDep}. 

\section{Experimental Measurements} 
\subsection{Experimental Setup and procedure}
\begin{figure} [htp]
			\centering
			\includegraphics[width=\linewidth]{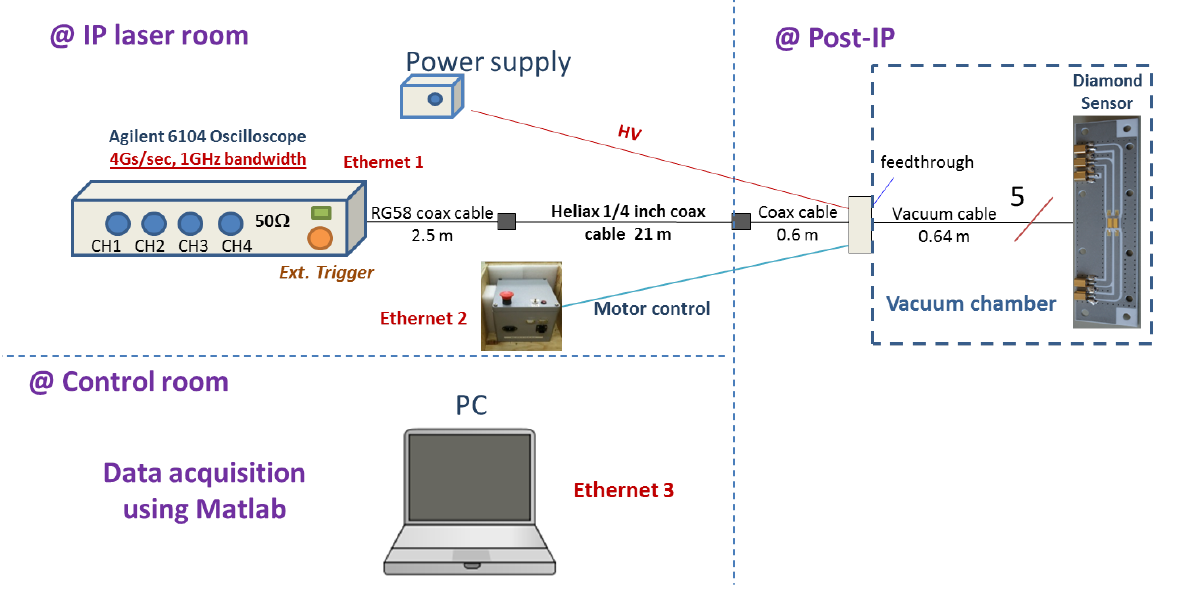} 
			\hspace{3pt}
			\caption{Layout of diamond sensor on ceramic PCB (right) and the data acquisition system (left)}
			\label{fig:DiamondLayout}
		\end{figure}

\indent Two detectors based on Chemical Vapor Deposition (CVD) single crystal diamond sensors have been built and installed after the IP. Each diamond sensor is 500 $\mu$m thick, with the metalization arranged in four strips,  two broad ones with the dimenssions of 1.5 mm $\times$ 4 mm and two narrow ones of 0.1 mm $\times$ 4 mm. The strips and related circuitry are mounted on a ceramic PCB and placed in vacuum. All the strips are biased at -400 V and connected to 50 $\Omega$ resistors by coaxial cables for signal readout by an oscilloscope, as shown in Fig.\ref{fig:DiamondLayout}. To suppress high frequency noise on the supplied bias voltage and to provide a sufficient reserve of charge for the largest signals, a low pass filter together with charging capacitors are mounted on the backside of the ceramic PCB~\cite{liu2016vacuum}. Since the DS are located behind a large bending magnet, the horizontal dispersion is close to 1 m for the "10$\times$1" optics.
	\begin{figure} [ht]
		\includegraphics[width=0.9\linewidth]{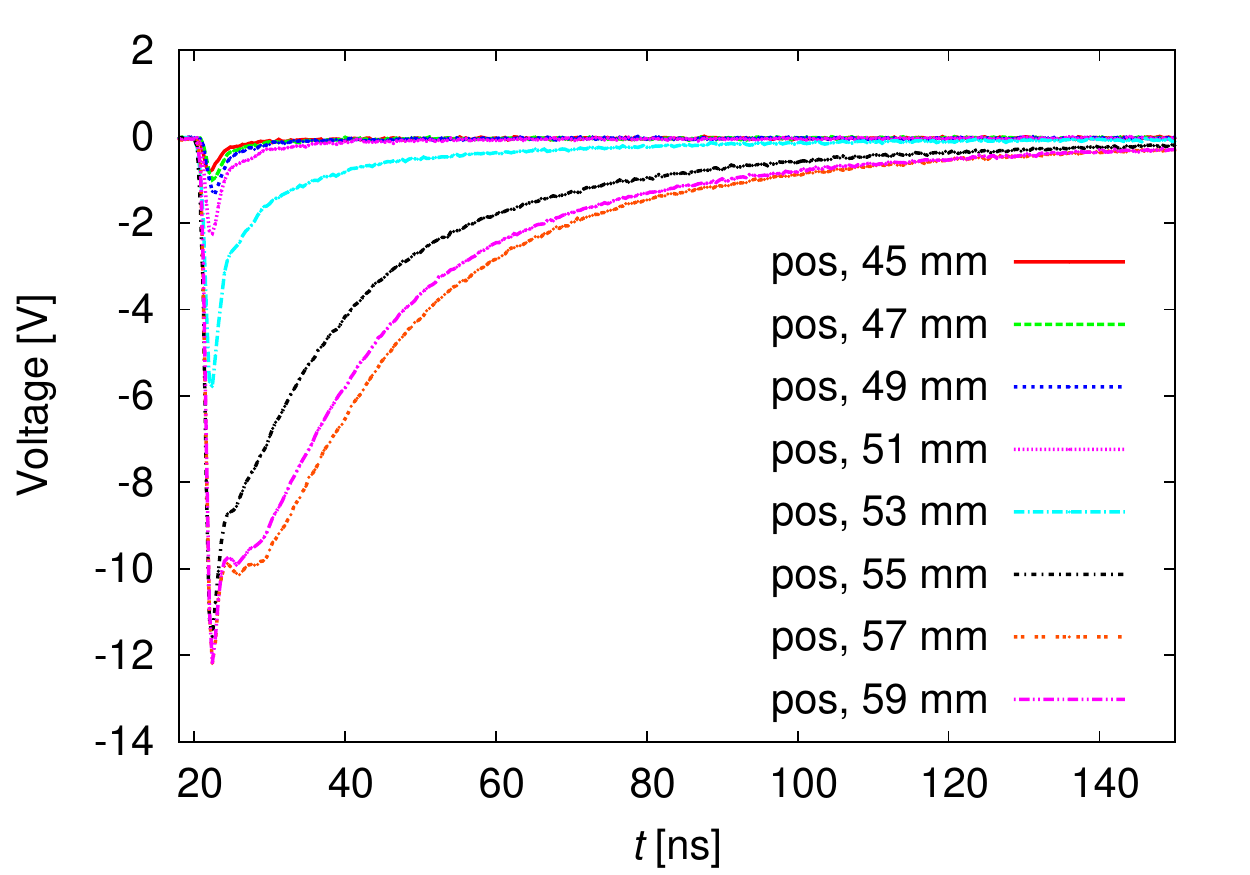}
		\put(-180, 135){(a)}\\
		\includegraphics[width=0.9\linewidth]{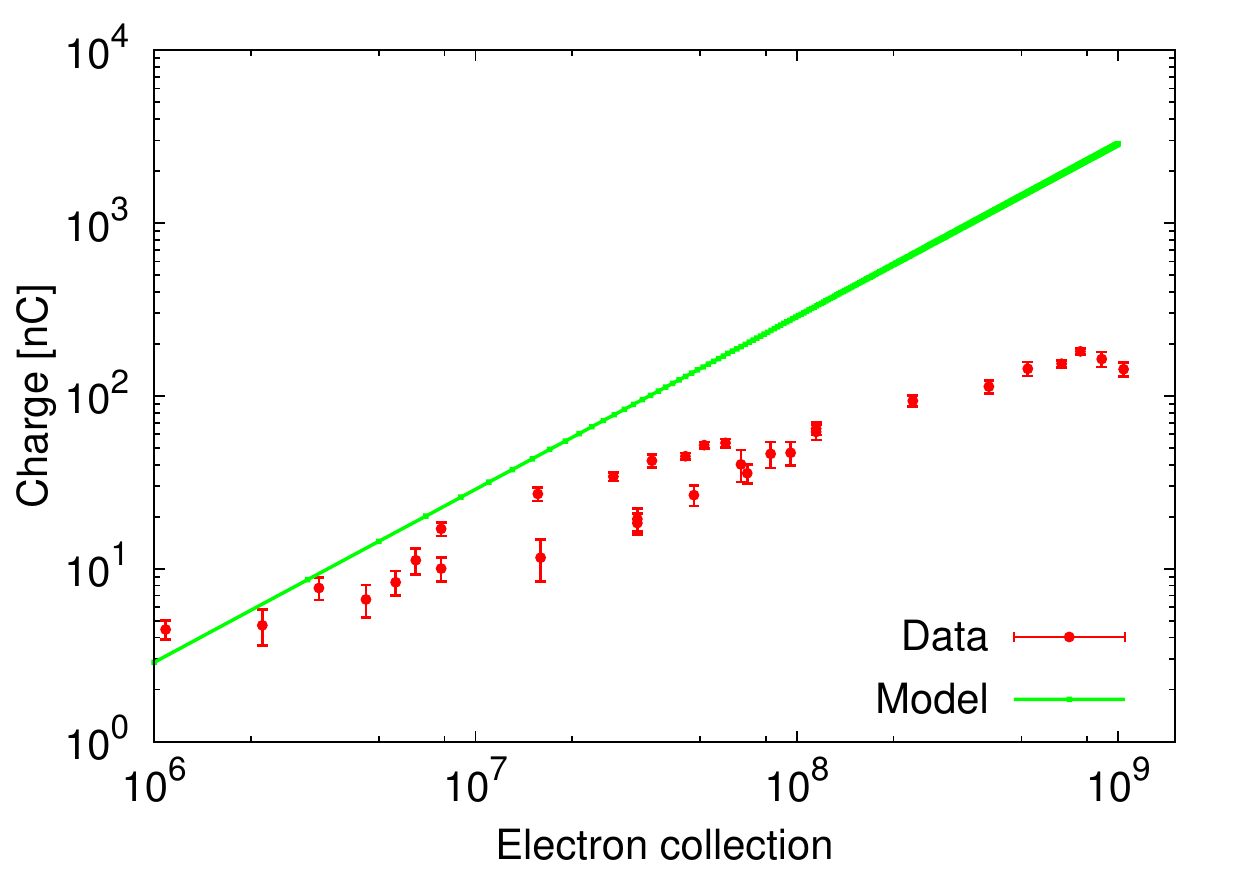}
		\put(-180, 135){(b)}
		\caption{Typical waveforms measured with the DS within the core region where beam center is at around 57 mm (a) and charge signal as a function of the quantity of collected electrons (b)}
		\label{fig:linearity_saturation}
	\end{figure}		
	
\indent The linear dynamic range of the diamond was demonstrated to be  10$^4$, with a lower limit of 10$^3$ electrons, which is determined by pickup noise induced by the passage of the beam in the vicinity, and a linear response up to 2$\times$10$^7$ electrons, which is limited by charge collection saturation effects in the diamond. Since a few thousand electrons is acceptable as background noise for the preliminary halo visualization, emphasis was put on the suppression of the saturation effect for the large signals. In the beam core region, readout becomes nonlinear and the waveform can be strongly distorted  both due to space charge inside the diamond crystal bulk and to the instantaneous voltage drop in the 50 $\Omega$ resistor, as shown in  Fig. ~\ref{fig:linearity_saturation} (a). The response of the output signal with respect to  the charge collected by the DS strip is shown in Fig.~ \ref{fig:linearity_saturation} (b). The number of electrons striking the diamond can be evaluated according to the beam intensity and transverse beam size, although this can involve some uncertainties, mainly from the determination of the beam size during scans and from instabilities at high intensity.
	 	\begin{figure} [h!]
	 		\includegraphics[width=0.9\linewidth]{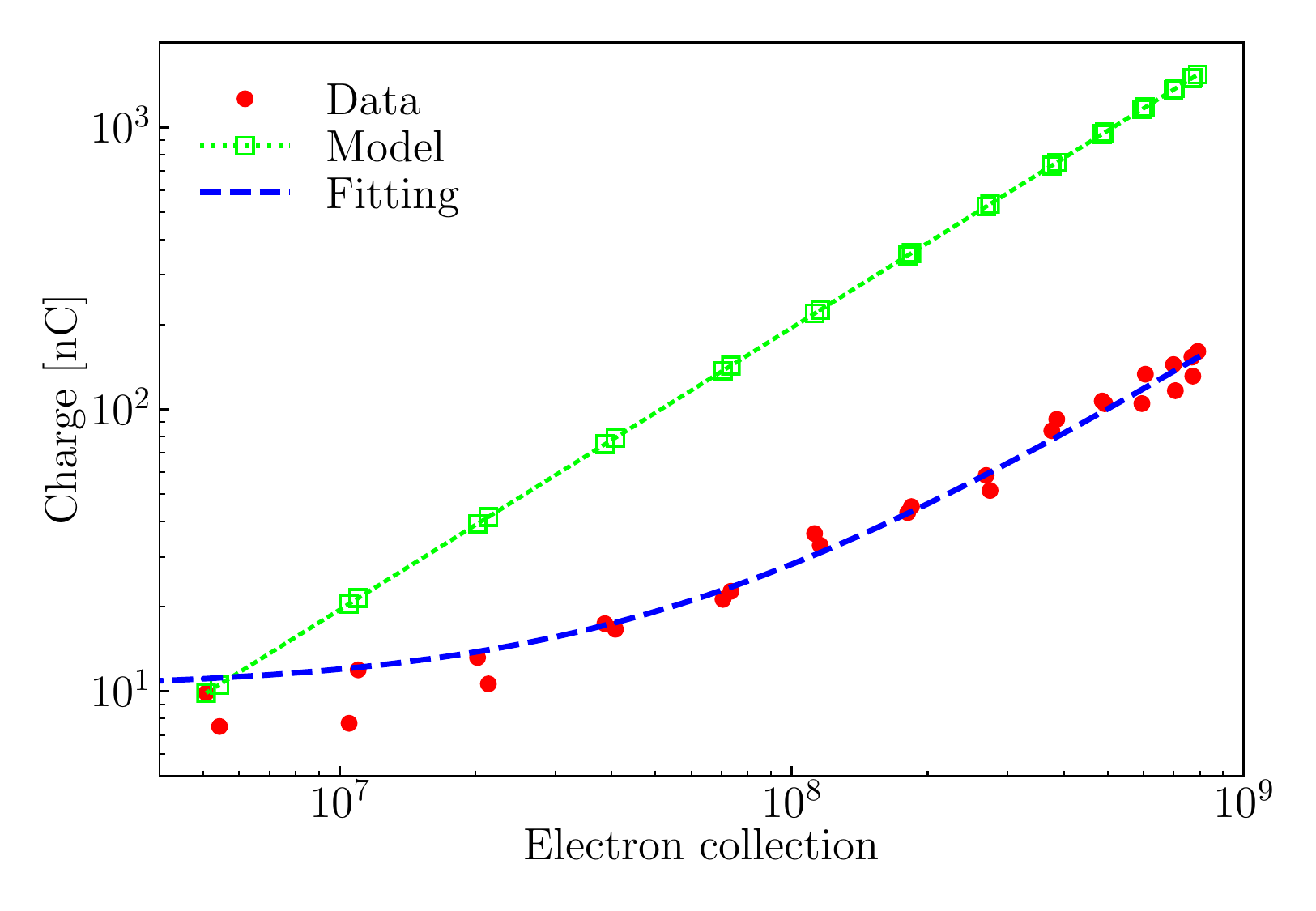}
	 		\put(-40, 130){(a)}\\
	 		\includegraphics[width=0.9\linewidth]{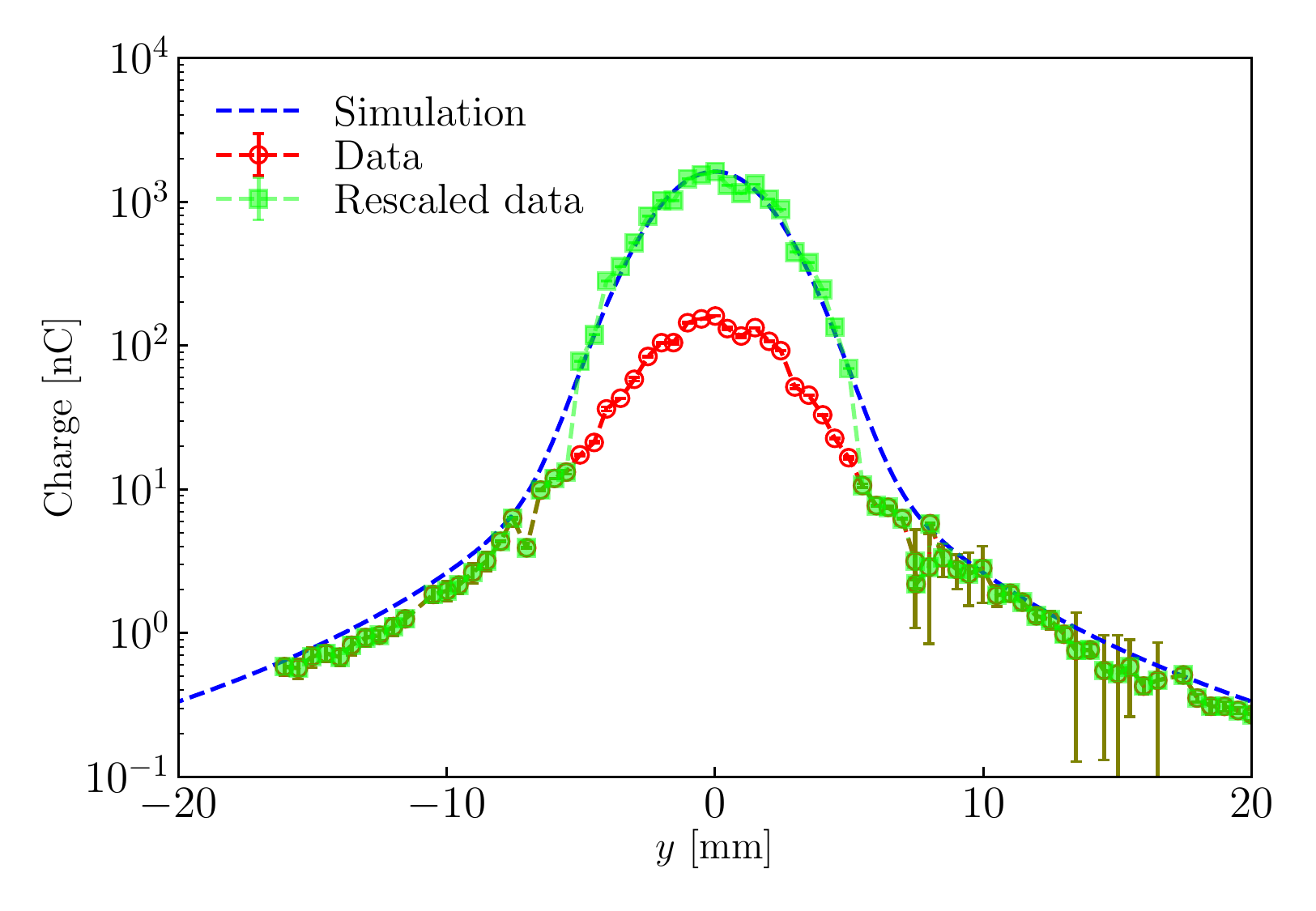}
	 		\put(-40, 130){(b)}
	 		\caption{Ideal and actual charge collections on DS strip as function of incident electron population for the evaluation of rescaling factor (a) and comparison of vertical beam profile before and after correction (b)}
	 		\label{fig:rescaling}
	 	\end{figure}	
	
\indent Rather than reconstructing the waveform based on the charge collection dynamics~\cite{kubytskyimodeling2015}, a "self-calibration" method was proposed to enable suitable correction of the core profile. The beam core distribution could be measured by the Wire Scanner (WS) located 2.89 m upstream and propagated to the DS to predict the number of electrons striking each strip according to its position with respect to the beam center. Subsequently, the charge $Q_{\text{exp}}$  which would be collected in the absence of saturation was computed based on the known electron hole pairs generation and charge collection efficiency measured at low incident charge~\cite{liu2016vacuum}. The rescaling factor $\kappa$ was then defined as the ratio of $Q_{\text{exp}}$ and the charge signal readout, and applied to rescale the DS data within beam core. After such rescaling based on "self-calibration", the linear dynamic range could be extended beyond 10$^{5}$ for the populations of collected electrons ranging from  $1\times10^3$ to more than $5\times10^8$. The corrected beam profile  is shown in Fig.~\ref{fig:rescaling}.
	 	\begin{figure*} [htpb]
	 		\centering
	 		\includegraphics[width=0.45\linewidth]{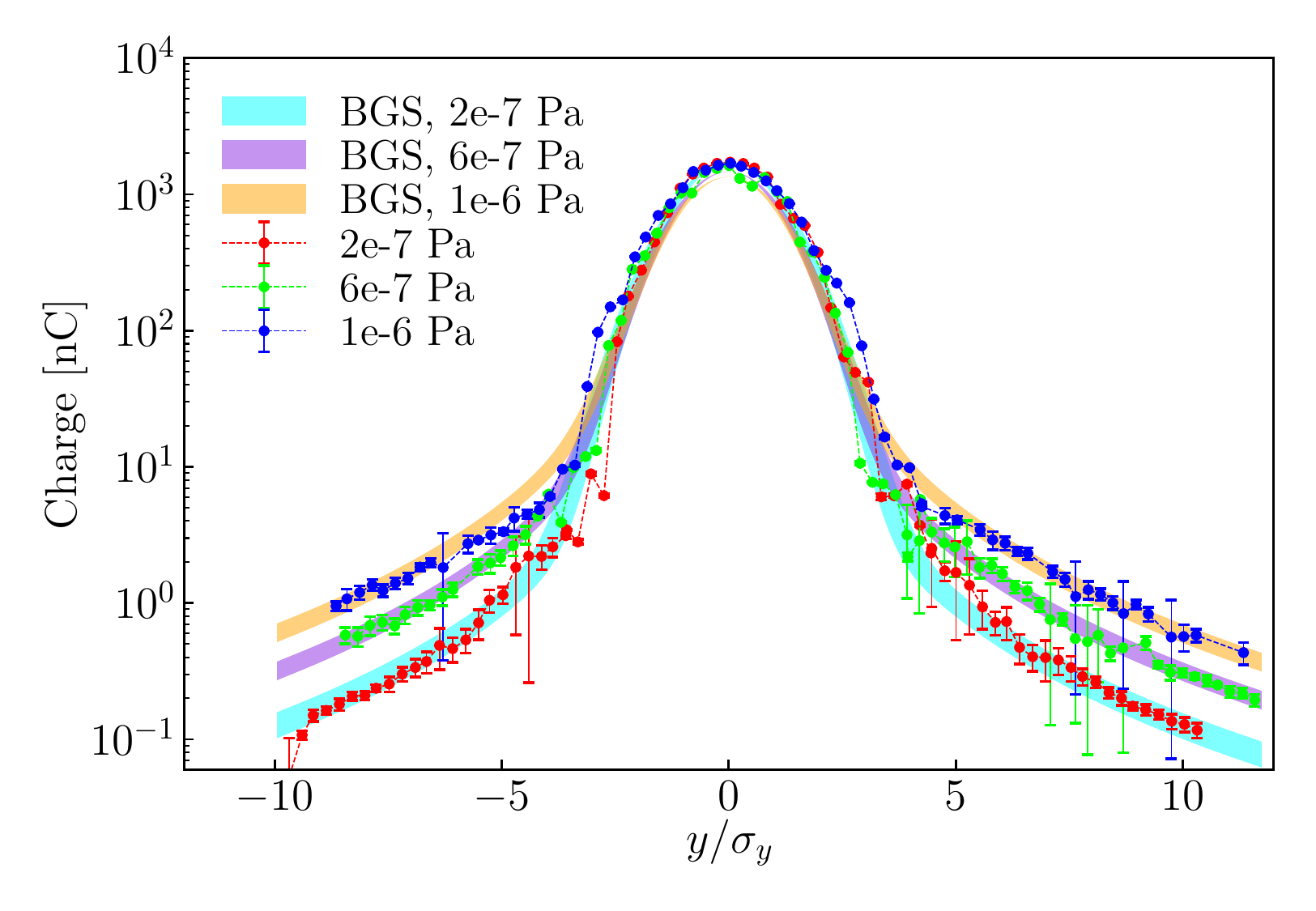}
	 		\put(-35, 130){(a)}
	 		\hspace{5pt}
	 		\includegraphics[width=0.45\linewidth]{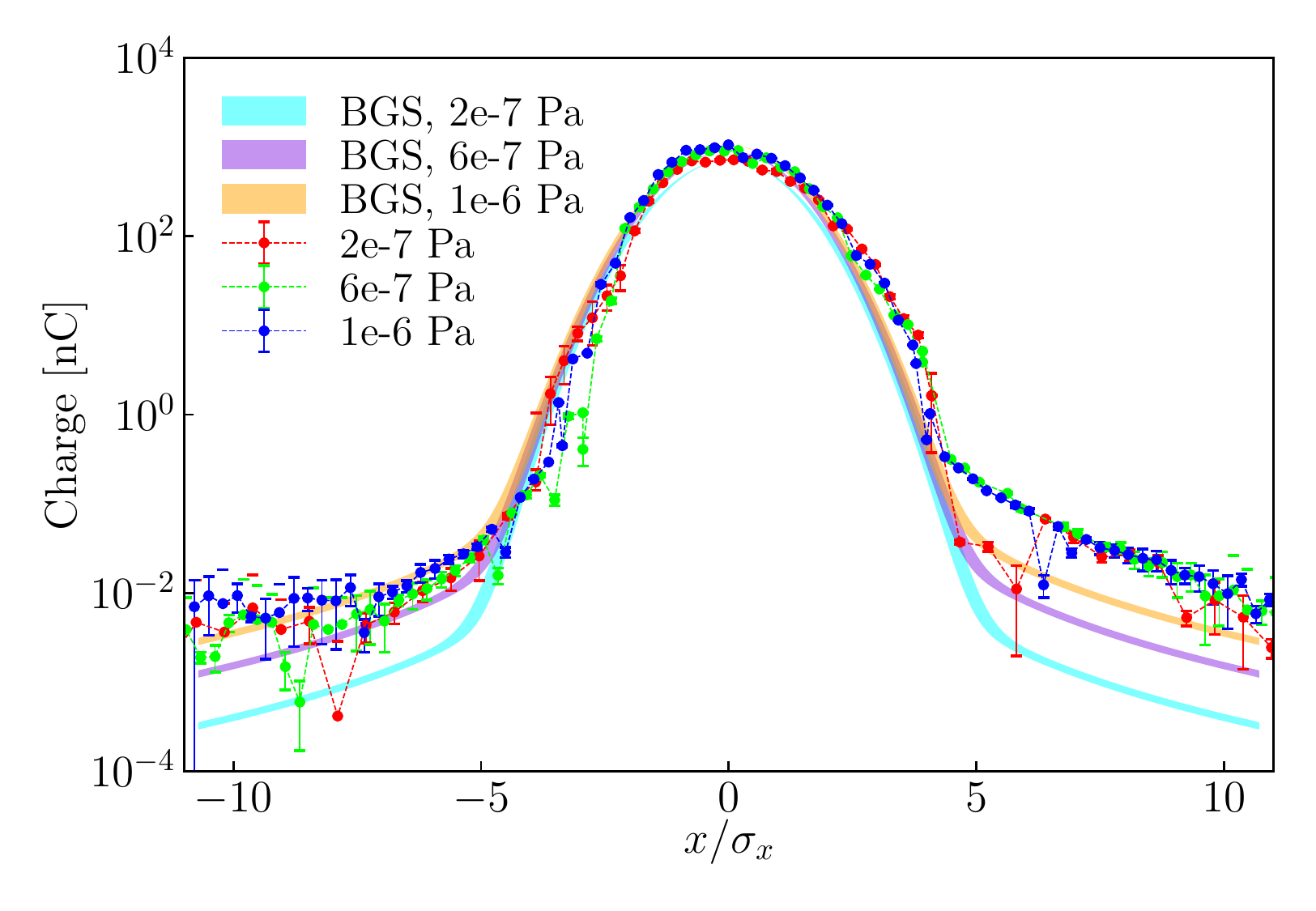}
	 		\put(-35, 130){(b)}
	 		\caption{Vertical (a) and horizontal (b) beam profiles normalized to beam core sizes for different vacuum pressures. The widths of the bands shown for the predictions from the BGS simulation represent the uncertainty of the beam size measurement}
	 		\label{fig:ds_hor_rescaled}
	 	\end{figure*} 	
\subsection{Transverse beam distribution}
\indent The transverse beam halo was measured using the DS for various vacuum pressures in the ATF damping ring. Beam intensity was stabilized at 3$\times10^{-9}$ $e$/pulse, and the residual gas pressure was increased by switching off SIPs in the arc sections and north straight section of the ATF damping ring. 

\indent Measured vertical beam halo distributions,  after implementing the rescaling corrections,  are consistent with our predictions from tracking simulations,  as shown in Fig.~\ref{fig:ds_hor_rescaled}(a). Moreover, the enhancement of vertical halo for the degraded vacuum pressures is clearly observed. Good agreement between simulations and experiments indicates that the dominant mechanism for vertical halo formation is elastic BGS in the ring.

\indent The measured horizontal beam distributions were also corrected using the described "self-calibration" method. The reconstructed beam profiles are higher than the predictions from BGS and asymmetrical distributions are observed, with more halo particles on the right side (high energy side), as shown in Fig.~\ref{fig:ds_hor_rescaled}(b). In addition, the evolution of the beam halo with the vacuum level was found to be negligible, which might be due to insufficient sensitivity, since the background noise level is around 0.01 nC. The DS being located in a high dispersion region after a large horizontal bending magnet ($\eta_x\approx1$ m),  potential non-Gaussian tails in the energy distribution of the beam may also play a role.
	 	
\section{Emittance Growth from Beam Gas Scattering}	
\indent 	Large angle beam-gas scattering events are rare but can induce large betatron oscillation amplitudes, which drive particles beyond the core and into the halo region. Meanwhile, small angle scattering events have higher probability and will act analogously to quantum excitation. They can  dilute the core particle distribution and cause emittance growth.

\begin{figure} [htpb]
				\centering
				 \includegraphics[width=0.97\linewidth]{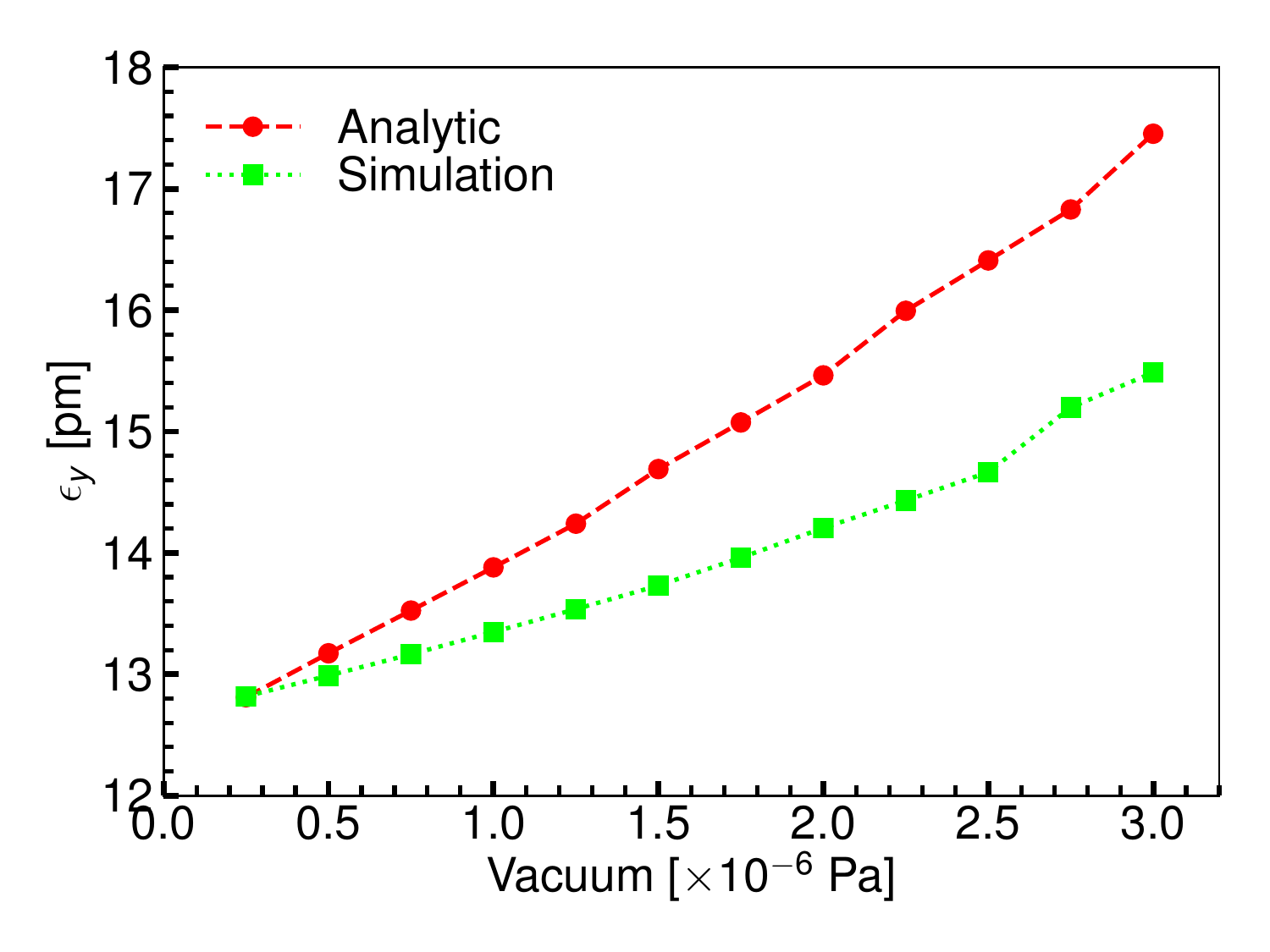} 
				\caption{Emittance growth as function of vacuum pressure predicted by Monte Carlo simulation and analytic calculation}
				\label{fig:simulationEmitGrowth0}
			\end{figure}

For typical vacuum pressures ($10^{-7}\sim10^{-6}$ Pa) at ATF, vertical emittance dilution is estimated with the beam distribution function derived in Sec. I and using the Monte Carlo simulation. We assume  that the worst vacuum pressure is 3$\times10^{-6}$ Pa and the equilibrium vertical emittance (without BGS and IBS) is 12.8 pm. This value is increased to 17.5 pm and 15.5 pm, as predicted by the analytic approximation and Monte Carlo simulation (see Fig. ~\ref{fig:simulationEmitGrowth0}), respectively. The difference between the two predictions is due to the approximation of the characteristic function integration in Eq.~\ref{eq:BGScrossapprox}.

\begin{figure} [htpb]
				\centering
				\includegraphics[width=0.97\linewidth]{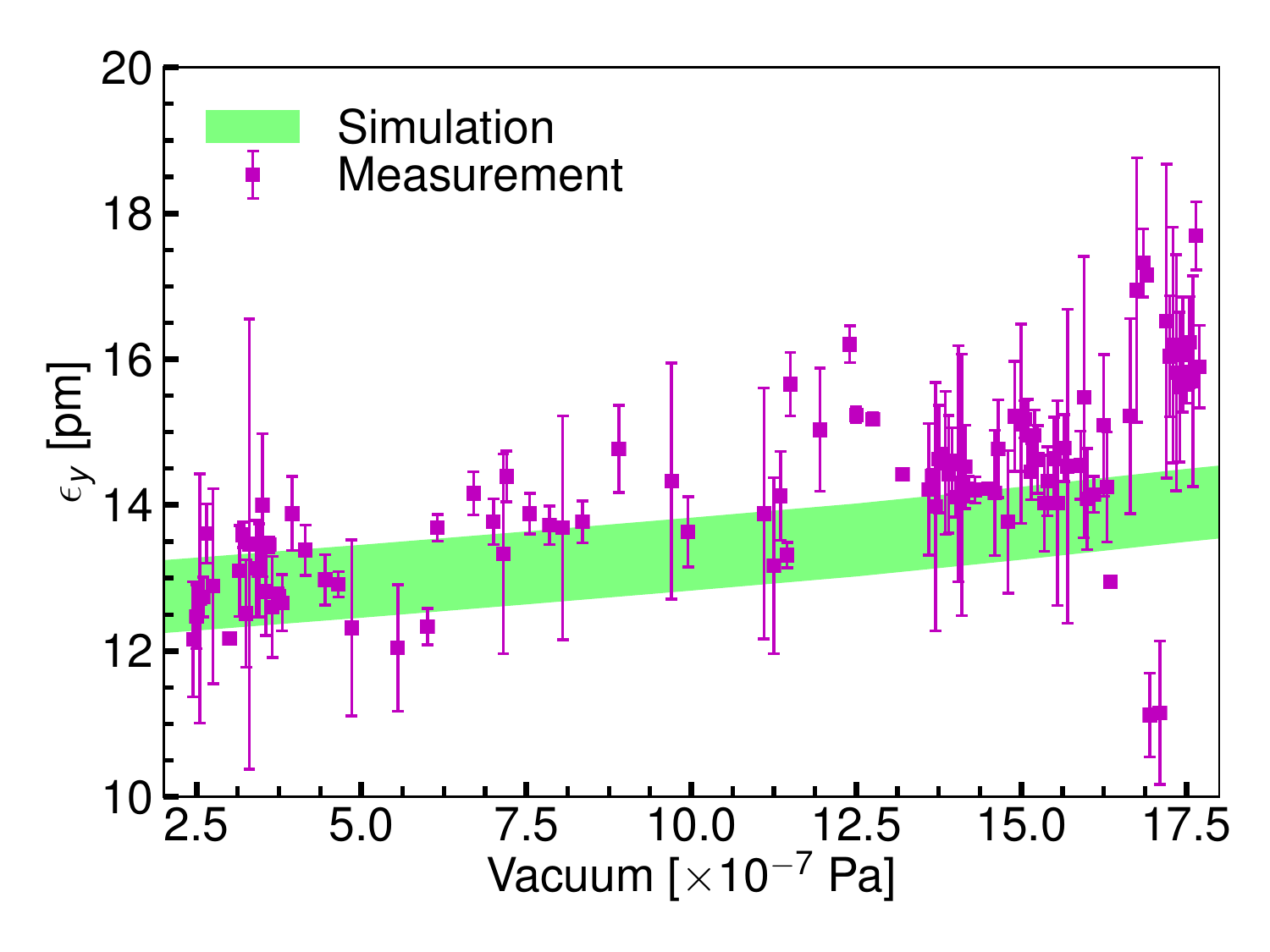}
				\put(-45, 147){(a)}\\				
				 \includegraphics[width=0.97\linewidth]{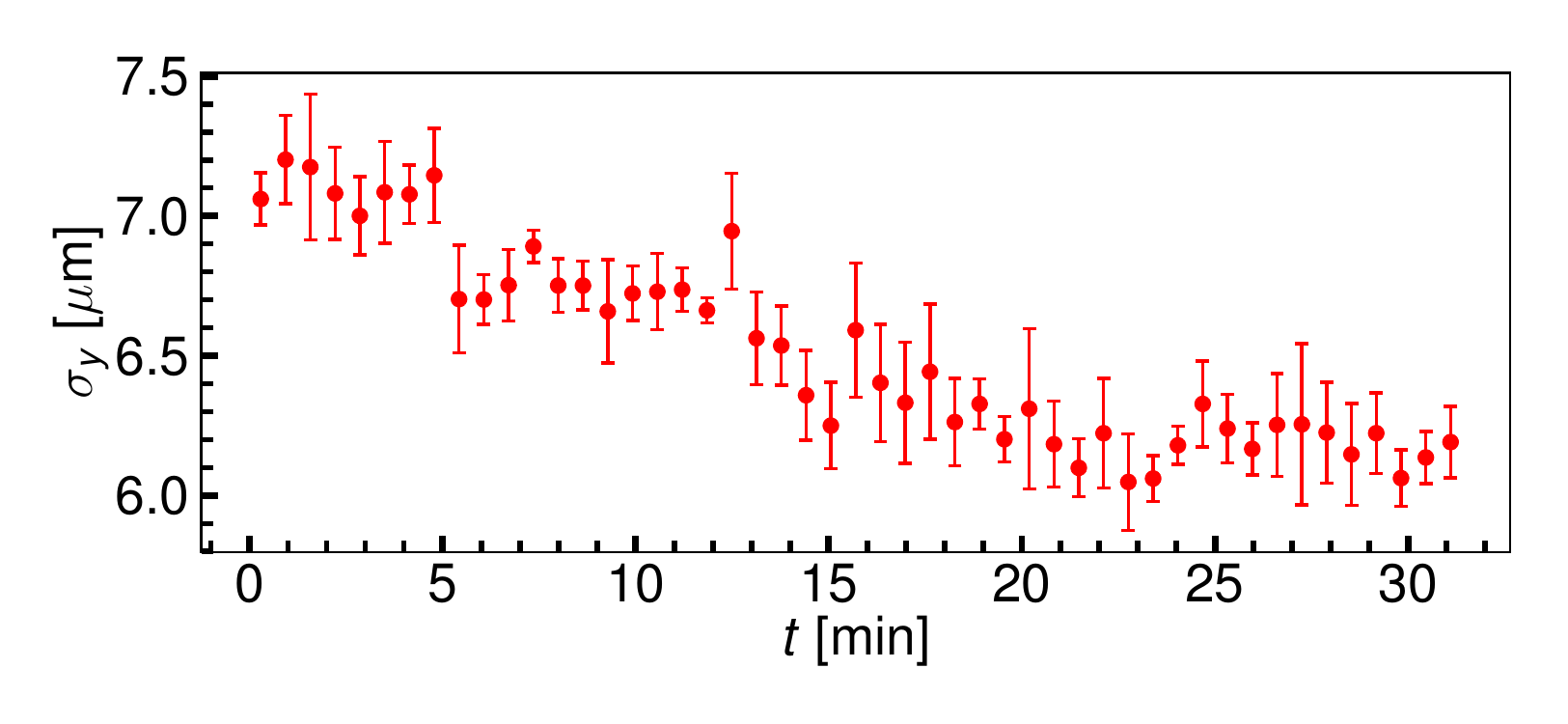}
				 \put(-45, 85){(b)}
				\caption{Evaluation of vertical emittance with respect to vacuum pressure of the ring (a) and evolution of beam size measured by XSR when all SIPs were reset at $t=0$ (b). Band width of the vertical emittance estimated by simulation is due to the $10\%$ uncertainty for the determination of the vertical emittance in absence of BGS}
				\label{fig:simulationEmitGrowth}
			\end{figure}

\indent To probe the above predictions, measurements of vertical emittance were performed for vacuum pressures ranging from 2.5$\times10^{-7}$ Pa to 1.75$\times10^{-6}$ Pa. Vertical emittance is evaluated from the beam size measured by an X-ray synchrotron radiation (XSR) monitor and the corresponding $\beta$ function~\citep{naito2012emittance}. The observed vertical emittance increases from 12.63 $\pm$0.46 pm to 16.02$\pm$0.98 pm , which is higher than the simulation result, see Fig.~\ref{fig:simulationEmitGrowth}(a). The difference might be caused by the uncertainty in the vacuum pressure measurement, the systematic error in the XSR monitor or some other physical process contributing to emittance growth~\citep{zimmermann1997studies}. Moreover, the vertical beam size monitored by the  XSR reduces from 7.02 $\mu$m to 6.2 $\mu$m when the vacuum pressure recovers from 1.75$\times10^{-6}$ Pa to 2.5$\times10^{-7}$ Pa, as shown in Fig.~\ref{fig:simulationEmitGrowth}(b). These pieces of evidence indicate that emittance growth due to BGS is also visible for typical vacuum pressure of $10^{-6}$ Pa and should be taken into account in the design of such a ring.
\section{Discussion and Conclusions}
\indent To explore the primary mechanism of halo formation at ATF, systematic analytical calculations, simulations and experimental measurements have been carried out. We applied formulas to approximate the beam distribution function in the presence of radiation damping, quantum excitation and BGS in the normalized coordinate system. The final simplified formalism, Eq.(\ref{eq:disfuc_bgs_final2}), is solely suitable for the estimation of beam halo, and not the beam core dilution.


\indent For an accurate prediction of the beam distribution distortion, a detailed Monte Carlo simulation was developed in the context of the SAD program. The actual COD and equilibrium beam parameters were approached by introducing local orbit bumps using steering magnets. We attempted to benchmark this simulation using the vacuum lifetime, which was found to be 83 and 87 minutes, from the two numerical methods, respectively, while the measured value was 16 minutes. The presence of additional horizontal beam halo, from sources other than BGS, and the reduction of the transverse acceptance due to nonlinear components may be the reasons for this difference.

\indent To extend the dynamic range of the diamond sensor detector used for the measurements, a rescaling scheme based on "self-calibration" was proposed and implemented to the DS data. After this rescaling correction, an effective dynamic range of $10^5$ can be achieved. Vertical and horizontal beam halo were measured for several vacuum pressures. For the vertical halo, good agreement between numerical estimations and experimental results for the different vacuum levels is observed. This clearly shows that the vertical halo is dominated by elastic BGS in the ring. On the other hand, the horizontal halo measured by the DS is higher than the BGS prediction and found to be asymmetric. The change in horizontal halo as a function of vacuum pressure is negligible. This shows that BGS has almost no influence on the horizontal halo distribution and other processes (eg. chromaticity, IBS and resonances) may play a more important role.

\indent Simulations and experimental observations of the vertical beam distribution clearly demonstrate that, for typical vacuum pressures in the ATF damping ring, halo generation and emittance growth due to BGS are both visible and significant. 

\indent Further studies of beam halo at ATF have been proposed, including the installation of a new OTR/YAG monitor at the dispersion-free region after the extraction from the ring, halo measurements for different kicker timings and optical focusing, and investigation of tails in the momentum distribution.

\begin{acknowledgments}
The authors express their gratitude to the ATF collaboration and to the staff and engineers of ATF.  One of us (R. Yang) would like to particularly thank K. Oide for the support and advice on the proper usage of the SAD software, offering suggestions and encouragement. This work was supported by the Chinese Scholarship Council, the Toshiko Yuasa France-Japan Particle Physics Laboratory (project A-RD-10), the France-China Particle Physics Laboratory (project DEV-IHEP-LAL-LC-CEPC) and the MSCA-RISE E-JADE project, funded by the European Commission under grant number 645479.

\end{acknowledgments}

\bibliography{aps_halobgs} 

\end{document}